\documentclass[aps,pra,reprint,floatfix,dvipsnames,bibnotes]{revtex4-1}

\usepackage[T1]{fontenc}
\usepackage{lmodern}
\usepackage{graphicx}
\usepackage{qcircuit}
\usepackage{braket} %To use quantum computing macros
\usepackage{amsmath,amsfonts,mathrsfs,bbm,bm,dsfont}
\usepackage{chemformula} %Chemical formulas.
\usepackage{hyperref}
\hypersetup{colorlinks=true,linkcolor=blue,citecolor=red,urlcolor=MidnightBlue}
\usepackage[capitalise]{cleveref}
\usepackage{microtype}
\usepackage{xspace}
\xspaceaddexceptions{]\}}

\usepackage[explicit]{titlesec}
\renewcommand{\thesubsection}{\Alph{subsection}}

\titleformat{\section}
  {\large\sffamily\bfseries}
  {}
  {0em}
  {#1}
  []
\titleformat{name=\section,numberless}
  {\large\sffamily\bfseries}
  {}
  {0em}
  {#1}
  []
\titleformat{\subsection}[runin]
  {\bfseries}
  {\thesubsection.}
  {0.5em}
  {#1. }
  []
\titlespacing*{\section}{0pc}{3ex \@plus4pt \@minus3pt}{5pt}
\titlespacing*{\subsection}{0pc}{2.5ex \@plus3pt \@minus2pt}{2pt}

\let\oldvec=\vec
\renewcommand{\vec}[1]{\mathbf{#1}}
\makeatletter
\g@addto@macro\bfseries{\boldmath}
\makeatother

\makeatletter
\newcommand{\customlabel}[2]{%
   \protected@write \@auxout {}{\string \newlabel {#1}{{#2}{\thepage}{#2}{#1}{}} }%
   \hypertarget{#1}{}
}
\makeatother

\newcommand{\kupccgsdDepth}{%
\begin{tabular}{lr}
Dipeptide     & Circuit depth \\
\hline
Alanine       & 1400          \\
Arginine      & 2900          \\
Asparagine    & 2000          \\
Aspartic Acid & 1950          \\
Cysteine      & 1600          \\
Glutamine     & 2300          \\
Glutamic Acid & 2250          \\
Glycine       & 1100          \\
Histidine     & 2500          \\
Isoleucine    & 2250          \\
Leucine       & 2250          \\
Lysine        & 2500          \\
Methionine    & 2150          \\
Phenylalanine & 2750          \\
Proline       & 1850          \\
Serine        & 1600          \\
Threonine     & 1900          \\
Tryptophan    & 3350          \\
Tyrosine      & 2900          \\
Valine        & 1950          \\
\end{tabular}
}

\newcommand{\allorbitalstable}{%
\begin{tabular}{lrrr}
Dipeptide     & STO-3G & cc-pVDZ & cc-pVTZ(-f)\\
\hline
Alanine       & 134    & 428     & 1598       \\
Arginine      & 282    & 904     & 3374       \\
Asparagine    & 198    & 616     & 2306       \\
Aspartic Acid & 194    & 596     & 2234       \\
Cysteine      & 170    & 500     & 1846       \\
Glutamine     & 226    & 712     & 2662       \\
Glutamic Acid & 222    & 692     & 2590       \\
Glycine       & 106    & 332     & 1242       \\
Histidine     & 242    & 748     & 2802       \\
Isoleucine    & 218    & 716     & 2666       \\
Leucine       & 218    & 716     & 2666       \\
Lysine        & 242    & 792     & 2950       \\
Methionine    & 226    & 692     & 2558       \\
Phenylalanine & 270    & 844     & 3158       \\
Proline       & 182    & 580     & 2166       \\
Serine        & 154    & 484     & 1810       \\
Threonine     & 182    & 580     & 2166       \\
Tryptophan    & 334    & 1032    & 3866       \\
Tyrosine      & 290    & 900     & 3370       \\
Valine        & 190    & 620     & 2310       \\
\end{tabular}
}

\newcommand{\activespacetable}{%
\begin{tabular}{lrr}
Dipeptide     & \hspace{5em}$M$ & \hspace{5em}$\eta$            \\
\hline
Alanine       & 112             &  64               \\
Arginine      & 236             & 132               \\
Asparagine    & 164             &  96               \\
Aspartic Acid & 160             &  96               \\
Cysteine      & 128             &  76               \\
Glutamine     & 188             & 108               \\
Glutamic Acid & 184             & 108               \\
Glycine       & 88              &  52               \\
Histidine     & 200             & 112               \\
Isoleucine    & 184             & 100               \\
Leucine       & 184             & 100               \\
Lysine        & 204             & 112               \\
Methionine    & 176             & 100               \\
Phenylalanine & 224             & 120               \\
Proline       & 152             &  84               \\
Serine        & 128             &  76               \\
Threonine     & 152             &  88               \\
Tryptophan    & 276             & 148               \\
Tyrosine      & 240             & 132               \\
Valine        & 160             &  88               \\
\end{tabular}
}

\usepackage[tooltip=true]{acro}

\newcommand{\defacro}[3]{\DeclareAcronym{#1}{
  short = #2,
  long = #3
}}

\defacro{aev}  {AEV}  {atomic environment vector}
\defacro{aiff} {AIFF} {\abinitio force field}
\defacro{ao}   {AO}   {atomic orbital}
\defacro{bch}  {BCH}  {Baker–Campbell–Hausdorff}
\defacro{bk}   {BK}   {Bravyi-Kitaev}
\defacro{cc}   {CC}   {coupled cluster}
\defacro{ccsdt}{CCSD(T)}{coupled-cluster singles and doubles with perturbative triples}
\defacro{cisd} {CISD} {CI Singles Doubles}
\defacro{ci}   {CI}   {configuration interaction}
\defacro{cnot} {CNOT} {Controlled-NOT}
\defacro{csf}  {CSF}  {configuration state function}
\defacro{fci}  {FCI}  {full configuration interaction}
\defacro{ff}   {FF}   {force-field}
\defacro{gto}  {GTO}  {Gaussian type orbital}
\defacro{hf}   {HF}   {Hartree-Fock}
\defacro{jw}   {JW}   {Jordan-Wigner}
\defacro{kupccgsd}{$k$-UpCCGSD}{$k$-Unitary paired Coupled Cluster Generalized Singles \& Doubles}
\defacro{lcao} {LCAO} {linear combination of atomic orbital}
\defacro{md}   {MD}   {molecular dynamics}
\defacro{mo}   {MO}   {molecular orbital}
\defacro{nisq} {NISQ} {noisy intermediate-scale quantum}
\defacro{nnp}  {NNP}  {neural network potential}
\defacro{nn}   {NN}   {neural network}
\defacro{noon} {NOON} {natural orbital occupation number}
\defacro{pes}  {PES}  {potential energy surface}
\defacro{qmmm} {QM/MM}{quantum mechanics/molecular mechanics}
\defacro{qpe}  {QPE}  {quantum phase estimation}
\defacro{qse}  {QSE}  {quantum subspace expansion}
\defacro{rdm}  {RDM}  {reduced density matrix}
\defacro{scf}  {SCF}  {self consistent field}
\defacro{si}   {SI}   {Supplementary Information}
\defacro{sto}  {STO}  {Slater type orbital}
\defacro{uccsd}{UCCSD}{UCC Singles \& Doubles}
\defacro{ts}   {TS}   {Trotter-Suzuki}
\defacro{ucc}  {UCC}  {unitary coupled cluster}
\defacro{vqe}  {VQE}  {variational quantum eigensolver}
\defacro{vqse} {VQSE} {virtual quantum subspace expansion}

\newcommand{\mytensor}[1]{\mathit{#1}}
\newcommand{\e}{\mathrm{e}}
\newcommand{\pmat}[1]{\begin{pmatrix}#1\end{pmatrix}}
\newcommand{\ii}{{i\mkern1mu}} % Imaginary i
\newcommand{\ketbra}[2]{|#1\rangle\!\langle#2|}
\newcommand{\bigO}{\ensuremath{\mathcal{O}}}
\newcommand{\abinitio}{\textit{ab initio}\xspace}
\newcommand{\Abinitio}{\textit{Ab initio}\xspace}
\newcommand{\hatT}{\hat{T}}

\date{\today}
\begin{document}

\title{High-Quality Protein Force Fields with Noisy Quantum Processors}

\author{Anurag Mishra}
\email{anurag@qulab.com}
\author{Alireza Shabani}
\affiliation{Qulab Inc., Los Angeles, California 90024, USA}

\begin{abstract}
  A central problem in biophysics and computational drug design is accurate modeling of
  biomolecules. The current molecular dynamics simulation methods can answer how a
  molecule inhibits a cancerous cell signaling pathway, or the role of protein misfolding
  in neurodegenerative diseases. However, the accuracy of current force fields
  (interaction potential) limits the reliability of computer simulations. Fundamentally a
  quantum chemistry problem, here we discuss optimizing force fields using scalable
  \abinitio quantum chemistry calculations on quantum computers and estimate the quantum
  resources required for this task. For a list of dipeptides for local parameterizations,
  we estimate the required number of qubits to be 1576 to 3808 with cc-pVTZ(-f) orbital
  basis and 88 to 276 with active space reduction. Using a linear depth ansatz with
  active-space reduction, we estimate a quantum circuit with a circuit depth of few
  thousands can be used to simulate these dipeptides.  The estimated number of 100s of
  qubits and a few thousand long circuit depth puts the pharmaceutical application of
  near-term quantum processors in a realistic perspective.
\end{abstract}
\maketitle

\setlength{\parskip}{2mm plus0.5mm minus0.5mm}

Structure and dynamics of proteins and other biomolecules determine their
functioning role in living organisms. How a protein folds shapes its structure and its
mechanistic interaction with other molecules in a cell. Therefore targeting biomolecules
with abnormal behavior is a prime therapeutic approach. Since the early success of protein
dynamics simulation ~\cite{levitt_1975_simproteinfold,warshel_1976_theoreticalstudies},
computer simulation of biomolecules has been a cornerstone of structural biology and drug
design \cite{durrant_2011_mddrugs}. The true dynamics of a protein system can be
completely described by solving the time dependent Schr\"{o}dinger equation to obtain the
motion of nuclei and electrons. This is a quantum problem that requires \abinitio quantum
chemistry methods. While great advances have been made in the field, solving systems
beyond $\sim50$ atoms remains an unfeasible task~\cite{guo_2018_dlpnoccsd}. For dynamical
systems, \abinitio techniques remain unfeasible even at a small size.

In order to make useful computational predictions at a large size, one can further
approximate the molecular system and assume that it is completely driven by Newtonian
mechanics. \Ac{md} applies classical mechanics to describe the dynamics and interactions
of molecules~\cite{cramer_2004_book}. Recent advances in free energy calculation has
turned \ac{md} simulation into a reliable tool for {\it{in-silico}} drug
discovery~\cite{wang_2015_accuratereliable,chipot_2007_freeenergy}. \Ac{md} finds wide
applicability in various fields, and has been used for calculating protein folding
kinetics~\cite{snow_2002_absolutecomparison}, for computing ligand-protein binding
energy~\cite{chakrabarti_2005_computationalprediction}, or deciphering CRISPR
mechanism~\cite{palermo_2019_crispr}. The multi-scale nature of these interactions, both
in time and space, along with the complexity of biomolecules, demands a full-atomistic
\ac{md} simulations. Over the years the accuracy of simulations have been significantly
improved. However, toward a full predictive power, the accuracy and speed of computer
simulation need further progress \cite{heo_2018_experimentalaccuracy}.

The accuracy of protein simulations relies on proper modeling of molecular
interactions. An \ac{md} trajectory captures motion of atoms nuclei where the dynamics is
governed by the energy potential shaped by the electronic cloud. A common approach
utilizes a classical potential function parametrized by local quantum chemistry
calculations or experimental fitting, basically a hybrid quantum-classical approach
\cite{hu_2008_freeenergyqmmm, warshel_1976_theoreticalstudies}.  More recent theories
suggest quantum perturbative methods ({\abinitio} force-fields) \cite{xu_2018_aiff} or
neural networks to replace the classical potential function
\cite{smith_2017_ani1,chmiela_2018_mlff}.

Here, we focus on the \ac{md} simulations driven by classical potentials and force-fields
and discuss how their quality can be improved by running {\abinitio} quantum chemistry
calculations on quantum computers. As an illustrative task, we focus our attention towards
optimizing protein force field parameters via \abinitio computations on a quantum
computer. This task, while not impossible, is prohibitively expensive to tackle with
classical computational techniques. Unlike a classical computer, the quantum resources to
perform an \abinitio computation scales linearly in the size of the
problem~\cite{lee_2019_kUpCCGSD,kivlichan_2018_quantumsimulation}. In this work, we first
briefly review force field models for approximate calculation of the dynamics of a
molecular system. We then discuss how \abinitio quantum chemistry simulations can be used
to improve protein force-field parameterization for more accurate \ac{md} simulations. We
provide an estimate for the quantum resources required for to perform this task on near
term quantum computers, and conclude with a discussion of other potential application of
quantum computing to the field of biophysics. Additional details are provided in the
Appendix. Our work shows that it is feasible to implement this illustrative task on the
\ac*{nisq} computing era.

%%%%%%%%%%%%%%%%%%%%%%%%%%%%%%%%%%%%%%%%%%%%%%%%%%%%%%%%%%%%%%%%%%%%%%%%%%%%%%%%%
\section{Theory}
\label{sec:theory}
%%%%%%%%%%%%%%%%%%%%%%%%%%%%%%%%%%%%%%%%%%%%%%%%%%%%%%%%%%%%%%%%%%%%%%%%%%%%%%%%%

\subsection*{Force Field}
\label{sec:ff-model}

The \ac{pes} of an atomic system describes its energy as a function of the chosen
coordinates. A \acf{ff} tries to approximate the \ac{pes} via a limited number of
classical terms. The accurate \ac{pes} depends on quantum mechanical effects, such as
exchange repulsion, which have no classical analog. Thus, a complete and accurate
description of the \ac{pes} of a molecule with only a finite number of classical
coordinates is an impossible task. Nevertheless, one can find a good approximation of this
energy surface near the equilibrium where the configuration of the system is not too far
from the stable configuration(s). A good \ac{ff} model tries to balance multiple goals:
it should use as little computational resources as possible to calculate the forces,
it should describe the \ac{pes} as accurately as possible and it should generalize
over any possible combinations of atoms and configurations. As one can imagine, these
goals are quite frequently in conflict with each other.

Common classical \ac{ff} models are described by the
potential~\cite{leach_2001_book,cramer_2004_book}
\begin{align}
  \label{eq:classical-ff}
  \begin{split}
  V  = \frac{1}{2}&\sum_{i>j} k_{ij} (r_{ij}-\bar{r}_{ij})^{2}
       + \frac{1}{2} \sum_{i} \tau_{i} (\theta_{i} - \bar{\theta}_{i})^{2} \\
       + \frac{1}{2} &\sum_{ni} V_{ni}\left(1+\cos(n\omega_{i}-\bar{\omega}_{i})\right)  \\
       + \phantom{\frac{1}{2}} &\sum_{i>j} \frac{q_{i}q_{j}}{r_{ij}}
       + \sum_{i>j} 4\epsilon_{ij}\left[
       {\left(\frac{\sigma_{ij}}{r_{ij}}\right)}^{12} -
       {\left(\frac{\sigma_{ij}}{r_{ij}}\right)}^{6}
       \right]
  \end{split}
\end{align}
where the first three terms describe the energy due to stretching, rotation and torsion of
the bonds respectively and the last two terms describe the Coulomb and the van der Waals
forces. $\bar{r},\bar{\theta}$ and $\bar{\omega}$ are the equilibrium bond distances, angles and
torsional angles respectively, $q_{i}$ are the charges on atoms and $\epsilon/\sigma$ are the van der
Waals constants. For better accuracy, one can augment the force fields with higher order
polynomials, such as a cubic term for bond stretching or an exponentially decaying
dispersion term. Or one can add many-body terms which describe the secondary effects of
two body interactions. However, computing special functions is more expensive than
evaluating polynomials and hence classical \ac{ff}s are usually limited to two-body
interaction terms and assume a simple polynomial form for most forces. Apart from the form
laid out in \cref{eq:classical-ff}, specialized \ac{ff}s also add additional coordinates
to better capture the behavior of a molecular simulation. For example, protein \ac{ff}s
include parameterization in terms of the protein backbone angles, etc. A good force field
should reproduce results obtained via known experiments and should be extensible so as to
provide useful predictions for other systems.

\subsection*{Protein force field parameterization}
\label{sec:protein-ff-params}

\Ac{md} simulations are usually deployed to study protein folding dynamics and to discover
stable and metastable conformational states~\cite{lee_2017_initioprotein}. The accuracy of
such \ac{md} simulations depend greatly on the quality of chosen force field, which itself
depends on the proper parameterization of various constants of the force field. Such
parameterization can be done at different levels, such as optimizing the entire force
field parameters simultaneously~\cite{wang_2014_buildingforce,
  robustelli_2018_developmdff} or by focusing on smaller set of parameters (say the
torsional terms) while keeping the rest of the terms fixed~\cite{best_2012_optcharmm}. In
either case, the \ac{md} simulations try to fit the computational results to known
reference data. The reference data can be either obtained experimentally or is often
generated from high quality \abinitio simulations.  Experimental reference data for such
parameterization are expensive to gather and designing proper experiments for novel
systems is a non-trivial task. Compared to gathering experimental data, \abinitio
simulations are cheaper to perform and can produce accurate energy surfaces for small
molecules to which the force fields might be fitted directly. There is a long history of
using \abinitio quantum chemistry simulations to improve protein force
field~\cite{ponder_2003_proteinff, lopes_2015_currentstatus}. In particular, the backbone
angle terms of the protein force field are often derived by fitting to the 2-D
Ramachandran plot obtained from dipeptide simulations~\cite{best_2012_optcharmm,
  kaminski_2002_developmentpolarizable}. The ability to perform such dipeptide \abinitio
simulations is hence critical to the task of improving the accuracy of protein force
fields~\cite{hermans_2011_aminoacid}. Due to computational complexity of high quality
\ac{cc} simulations, current simulations of such dipeptides are often performed at a lower
level of theory~\cite{echenique_2008_efficientmodel, mironov_2019_alaninedipep}.

\Abinitio quantum chemistry simulations try to simulate the behavior of a quantum system,
viz.\ a small collection of atoms. Unsurprisingly, the resource requirement to do an exact
\abinitio calculation scales exponentially with the number of atoms in the system. A
universal quantum computer can simulate any quantum system with at-most a polynomial
overhead~\cite{lloyd_1996_universalquantum}. It is then reasonable to argue that a quantum
computer which includes quantum effects natively in its hardware should be used to perform
such \abinitio quantum chemistry calculations~\cite{feynman_1982_simulatingphysics}.  In
particular, a quantum computer can be used to simulate a molecule's \ac{pes}, and
hence can be used to perform \abinitio quantum chemistry calculations. \Ac*{qpe} can
provide comparable accuracy to \ac{fci} methods~\cite{abrams_1999_qpe} and \ac*{vqe}
methods should produce results comparable to coupled cluster
theory~\cite{takeshita_2019_vqse,kuhn_2019_accuracyresource}.  As always, care must be
taken to translate \abinitio results obtained in gas phase~\cite{rizzo_1999_oplsallatom}
before they are translated into protein \ac{ff} parameters which will be applied mostly to
aqueous phase.

\subsection*{\Abinitio Quantum Chemistry on Quantum Computer}
\label{sec:qc-ai}

Solving the Schr\"{o}dinger equation of a molecular Hamiltonian is an especially hard
problem. In \abinitio quantum chemistry methods, this problem is solved iteratively. We
build an approximate solution by neglecting some aspect of the Hamiltonian and this
solution is used as a starting point for the next iteration where a few more terms of the
Hamiltonian are added to the calculation. \Abinitio methods can be divided into two
groups; the \ac{hf} method~\cite{hartree_1928_wavemech1,*hartree_1928_wavemech2,
  *fock_1930_nherungsmethodezur,*slater_1930_notehartrees} which attempts to find the mean
field solution of the problem and post-\ac{hf} methods which attempt to systematically
improve on the \ac{hf} solution. We describe the details of \ac{hf} method in
\cref{sec:hf} and focus our discussion on the post-\ac{hf} methods. The post-\ac{hf}
methods become particularly easy to analyze in second quantized formulation of the quantum
Hamiltonian~\cite{helgaker_2014_elecstructure}:
\begin{equation}
  \label{eq:2quantHam}
  \hat{H} = \sum_{ij} h_{ij}a^{\dagger}_{i}a_{j} + \sum_{ijkl} V_{ijkl} a^{\dagger}_{i}a^{\dagger}_{j}a_{l}a_{k}
\end{equation}
where $a^{\dagger}_{i}$ ($a_{i}$) are the creation (annihilation) operators that add (remove) an
electron to orbital $i$ and the terms $h_{ij}$ and $V_{ijkl}$ describe the kinetic and
potential energy of the Hamiltonian. We provide a detailed analysis of the second
quantization method in \cref{sec:2quant}.

The \acf{cc} method~\cite{bartlett_2007_cctheory} is one post-\ac{hf} method which is
widely used for computing accurate properties of small molecules. The \ac{cc} method
starts with a reference wave function (usually the \ac{hf} wave function) that describes a
list of orbitals, of which the low energy orbitals are occupied while the high energy
orbitals remain empty. The \ac{cc} method constructs an exponential ansatz (a trial wave
function) by exciting some electrons from occupied orbitals to empty orbitals which can be
mathematically written as
\begin{equation}
  \label{eq:ccwave}
  \begin{aligned}
    \ket{\Psi_{\rm CC}} &= \exp\left(\hat{T}_{1} + \hat{T}_{2} + \ldots \right)\ket{\Psi_{\rm HF}}\ , \\
    \hat{T}_{1} &= \sum_{i,a}t^{a}_{i} a^{\dagger}_{a}a_{i} \ , \\
    \hat{T}_{2} &= \sum_{ij,ab} t^{ab}_{ij}a^{\dagger}_{a}a^{\dagger}_{b}a_{j}a_{i}\ , \text{\ etc.}
  \end{aligned}
\end{equation}
where the indices $i,j,\ldots$ run over the occupied levels, $a,b,\ldots$ run over the unoccupied
level and $\ket{\Psi_{\rm HF}}$ is the reference wave function obtained from the \ac{hf}
method. Different excitations are given different coefficients
($t_{i}^{a}, t_{ij}^{ab}, \ldots$) and these coefficients are optimized to give the best
solution. The \ac{cc} equations are usually solved via a projective method (see
\cref{sec:posthf}) which does not preserve the variational nature of the ansatz. Thus, the
computed energy is no longer an upper bound on the true ground state energy of the system.

The \ac{ucc} method is a modification of the \ac{cc} method where the ansatz maintains
its variational nature. This is achieved by considering both the excitation of electrons
from occupied to unoccupied orbitals and their relaxation back to the original orbitals,
\begin{equation}
  \label{eq:uccwave}
  \ket{\Psi_{\rm UCC}} = \exp\left( \hat{T}_{1} - \hat{T}_{1}^{\dagger} + \hat{T}_{2} - \hat{T}_{2}^{\dagger} + \ldots \right)
  \ket{\Psi_{\rm HF}}
\end{equation}
This maintains the variational nature of the system as $\exp(\hat{T}-\hat{T}^{\dagger})$ is a
unitary operator. The \ac{ucc} ansatz of \cref{eq:uccwave} can be efficiently prepared by
a quantum computer, and the ground state energy can be found by minimizing the expectation
value $E = \braket{\Psi_{\rm UCC}|H|\Psi_{\rm UCC}}$.

In the \ac{vqe} algorithm, a quantum
computer is used to prepare the ansatz and a classical optimizer optimizes the parameters
of the ansatz~\cite{peruzzo_2014_vqe,omalley_2016_scalablevqe,mcclean_2016_theoryvqe}. The
energy found via the \ac{vqe} algorithm remains an upper bound to the true ground state
energy of the molecular Hamiltonian and is hopefully more accurate for complex systems.
Since the method is variational, an error in preparation of \ac{ucc} ansatz
(\cref{eq:uccwave}) reflects as a slightly different optimal value of its
coefficients. Thus, the \ac{vqe} method is especially suitable for the noisy computers in
the \ac{nisq}~\cite{preskill_2018_nisq} computing era. As a downside, depending on the \ac{pes} of
the molecule and the quality of the classical optimizer, \ac{vqe} might take a long
time to find the minimum energy or get stuck in a local minima. In general, the \ac{vqe} algorithm can be
used with any ansatz that can be prepared by applying a unitary operator on the reference
wave function,
\begin{equation}
  \label{eq:vqe-unitary-ansatz}
  \ket{\Psi} = U_{\text{ansatz}} \ket{\Psi_{\rm HF}}
\end{equation}
Different types of ansatz might be chosen on the basis of hardware connectivity, the ease of
preparation, rate of convergence, etc. Thus, some of the above mentioned downsides may be
alleviated by a good choice of ansatz.

On current generation of quantum computers, small molecules such as
\ch{H2}~\cite{omalley_2016_scalablevqe,kandala_2017_hardwarevqe,hempel_2018_quantumchemistry},
\ch{HeH+}~\cite{peruzzo_2014_vqe,shen_2017_hehvqe},
\ch{LiH}~\cite{kandala_2017_hardwarevqe} and \ch{BeH2}~\cite{kandala_2017_hardwarevqe}
have been simulated to good accuracy by utilizing up to six qubits. The \ac{vqe} algorithm
has also been applied to compute the energy of an atomic
nucleus~\cite{dumitrescu_2018_cloudquantum}. We provide further details of the \ac{vqe}
algorithm in \cref{sec:vqe}.

%%%%%%%%%%%%%%%%%%%%%%%%%%%%%%%%%%%%%%%%%%%%%%%%%%%%%%%%%%%%%%%%%%%%%%%%%%%%%%%%%
\section{Results}
\label{sec:results}
%%%%%%%%%%%%%%%%%%%%%%%%%%%%%%%%%%%%%%%%%%%%%%%%%%%%%%%%%%%%%%%%%%%%%%%%%%%%%%%%%

\subsection*{Resource Estimate for FF Parameterization}
\label{sec:resource-estimation}

Like classical algorithms, it is essential to estimate the resources required for
implementing a quantum algorithm. Such estimates help us in understanding the practical
application of the given quantum algorithm and guide further optimizations. In quantum
computing, the number of qubits and the number of gates are the two physical resources
required to implement a quantum algorithm.

A qubit serves as the fundamental unit of information in quantum
computing~\cite{nielsen_chuang_book}. Classical algorithms are often
constrained by the amount of available memory, which is the number of bits required to
represent and process the problem. Similarly, quantum algorithms are constrained by the
number of qubits required to implement them on a quantum computer. In this aspect, the
number of qubits play a role similar to the memory size of a classical computer. Just like
a classical computer, the qubits can either hold the information about the problem
variable or they might hold temporary or ancillary information required in the course of
computation. A qubit can exist as a superposition of state $0$ and $1$ and can be
represented by a two element column vector or in the Dirac's bra-ket
notation~\cite{dirac_1981_qmbook} as
\begin{equation}
  \label{eq:qubit}
  \ket{\psi} = \alpha\ket{0} + \beta\ket{1} \equiv
  \begin{pmatrix} \alpha \\ \beta \end{pmatrix}
\end{equation}
where $\alpha$ and $\beta$ are complex numbers. When measured, the qubit in state
$\ket{\psi}$ will report value $1$ with probability $|\alpha|^{2}$ and the value $0$ with
probability $|\beta|^{2}$. Since $0$ and $1$ are the only two possible values, the state
$\ket{\psi}$ must be normalized such that $|\alpha|^{2}+|\beta|^{2}=1$. Two qubits can be in
four possible states, $00$, $01$, $10$ and $11$ and can be represented by a $4$ element
column vector. In general, a $n$-qubit state can be represented by $2^{n}$ element column
vector.
\begin{table}[t]
\centering
\allorbitalstable{}
\caption{\textbf{Qubit estimation.} The number of qubits required to simulate dipeptides
  in various basis sets without an active space approximation. All orbitals, including the
  core orbitals, are included in the \ac{vqe} computation.}
\label{tab:1}
\end{table}
\begin{table}[t]
\centering
\activespacetable{}
\caption{\textbf{Qubit estimation with active space approximation.} The number of qubits
  $M$ required to simulate dipeptides with $\eta$ valence electrons in with full reaction
  space approximation. The number of qubits $M$ are independent of the choice of the basis
  set, and depend only on the number of valence orbitals of each atom.}
\label{tab:2}
\end{table}

A quantum gate represents an action on the state of one or many quantum qubits. A quantum
gate acting on $n$-qubits can be represented by a $2^{n}\times2^{n}$ unitary matrix $U$
\begin{equation}
  \label{eq:qgates}
  U =
  \begin{pmatrix}
    u_{11} & u_{12} & \ldots & u_{1N} \\
    u_{11} & u_{12} & \ldots & u_{1N} \\
    \vdots      & \vdots      & \ddots &  \vdots \\
    u_{N1} & u_{N2} & \ldots & u_{NN}
  \end{pmatrix}
\end{equation}
where $N=2^{n}$ and $U^{\dagger}U=\mathds{1}_{N\times N}$. Each quantum gate takes some time to
complete its action which is called the gate operation time. If all gates act one after
the other, the time required for the quantum algorithm is simply the sum of the operation
time of each gate. Quite frequently, a quantum algorithm can split into a series of steps
such that gates in each of these steps can be applied simultaneously. The number of such
steps is called the circuit depth of the circuit. In this case, the time to complete the
algorithm is equal to the circuit depth times the operation time of the slowest gate. The
circuit depth is a measure of the time complexity of a quantum algorithm, but the number
of gates is a good measure of the quantum resource requirement since it dictates the
physical complexity of the hardware.

Here, we shall focus on the number of qubits and gates, and circuit depth required to
implement quantum \abinitio simulations for protein force field parameterization.

\subsection*{Qubit count}
\label{sec:qubit-count}

To get a qubit estimate for optimizing protein \acf{ff} parameters, we first start with an
estimate for full quantum computation of the electronic wave function of dipeptide
molecules. The number of qubits required to simulate a system is double the number of
basis functions, one for each spin orbital function (See
Methods, \ref{method:jw-transform}). The basis functions are chosen from a basis set during the
\ac{hf} optimization. We provide further details of basis sets in the
\cref{sec:basis-sets}. In \cref{tab:1}, we show the number of qubits required to simulate
dipeptide molecules in various minimal and split-$\zeta$ basis sets. This is a worse case
estimate and the structure of the molecular Hamiltonian can suggest several optimizations
that can reduce this resource requirement.
\begin{figure*}
  \centering
  \includegraphics{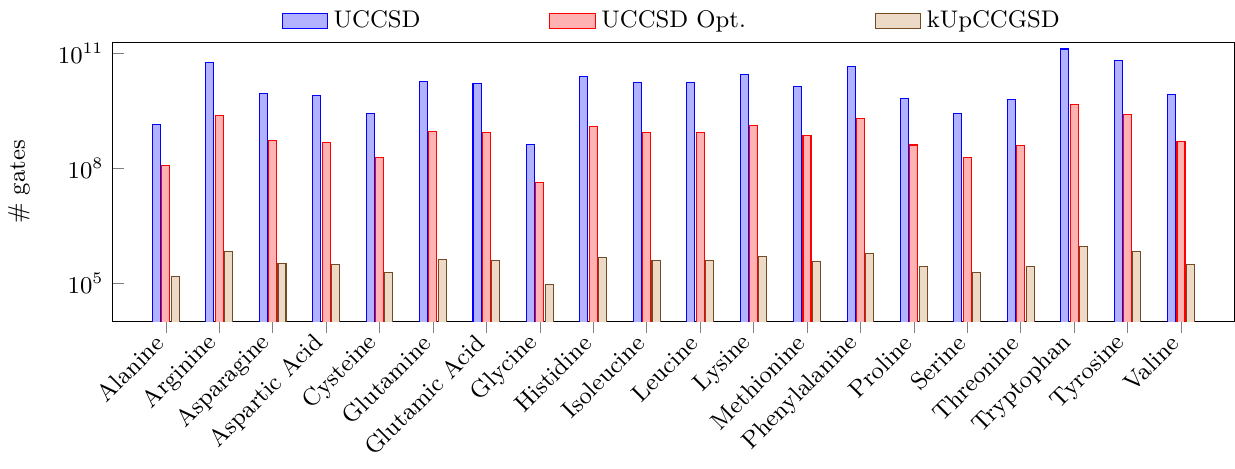}
  \caption{\textbf{Number of two-qubit gates required to implement \acs*{vqe} with different
      ansatz.} For a molecule of $\eta$ electrons and $M$ spin orbitals, the UCCSD ansatz in
    the \acs*{jw} encoding requires $O(M(M-\eta)^{2}\eta^{2})$
    gates~\cite{whitfield_2011_hamqpe}. The bar plot with legend UCCSD shows the number of
    two-qubit gates required by various dipeptides. However, clever placement of terms can
    cancel the overhead of \acs*{jw} encoding~\cite{hastings_2015_improvingquantum} which
    reduces the number of gates by $O(M)$. We term this as the UCCSD Optimized method. The
    largest gains are made by switching to a linear scaling ansatz, such as the
    \acs*{kupccgsd}~\cite{lee_2019_kUpCCGSD} ansatz. We show the $k=1$ case.}
  \label{fig:1}
\end{figure*}

The simplest way to reduce the qubit requirement is to reduce the number of basis
functions. Much of the chemical behavior of atoms can be solely attributed to the
arrangement of their valence electrons, the inner electrons don't participate in bond
formation and high energy orbitals far beyond the molecular energy scale will never be
occupied. This leaves the valence orbitals and few virtual orbitals next to the valence
orbitals as the only important orbitals for \abinitio simulations. Such methods, where one
only keeps certain orbitals in the post-\ac{hf} computation, are called active space
methods. Different active space methods differ in their choice of such orbitals.  The full
reaction space~\cite{ruedenberg_1982_fors} or the minimal molecular basis method selects
an active space that contains the same number of valence molecular orbitals as one
constructed from a minimal atomic basis set. This space contains the bonding, nonbonding
and antibonding orbitals of the molecule, and ignore the core orbitals. If the minimal
atomic basis set has $n$ valence functions, then the active space is constructed from $n$
low energy molecular orbitals. This approximation works reasonably well for post-\ac{hf}
methods that consider configurations representing states excited from the reference
states~\cite{townsend_2019_posthf, kowalski_2005_activaspacecc}. With this approximation,
\cref{tab:2} lists the number of qubits required to simulate various dipeptides on a
quantum computer with full reaction space. The number of qubits required for such
computation is independent of the chosen basis set and is equal to the number of qubits
required by a minimal atomic basis set, but the values of terms in
\cref{eq:2quantHam} will depend on the chosen basis set. The active space method greatly
reduces the number of qubits required to do useful computation, potentially at the cost of
less accurate results. In particular, an active space restricted to full reaction space
might not capture dynamical corrections to the \ac{hf}
energy~\cite{mok_1996_ndce,krylov_1998_vodcc}.

The qubit requirement can be reduced even without losing any information about the
system. This follows from a simple dimensionality analysis argument. Given $M$ spin
orbitals with $n$ electron, the possible number of valid many-electron configurations
scale as $\sim {M \choose n}$ which is smaller than the dimension of Hilbert
space~\cite{nielsen_chuang_book} described by $M$ qubits, $2^{M}$. Thus, the number of
qubits $Q$ required to describe the system satisfy the relationship
\begin{equation}
  \label{eq:combi-argument}
  {M \choose n} \leq 2^{Q} \leq 2^{M} \ .
\end{equation}
Finding the best encoding with such consideration remains an active topic of
research~\cite{takeshita_2019_vqse}. The qubit tapering
method~\cite{bravyi_2017_taperingqubits} utilizes the symmetries of the Hamiltonian to
reduce the number of required qubits. Since the number of electrons and the total spin of
the molecule is fixed, qubit tapering can always remove $2$ qubits. Due to lack of any
other spatial symmetry, we were not able to remove any additional qubits. We describe our
work in \cref{sec:qubit-tapering}.

\subsection*{Gate count}
\label{sec:gate-count}

The number of gates required to simulate the molecular Hamiltonian via the \ac{vqe}
algorithm depends on the choice of the ansatz. Here we discuss two
different ansatz, the \ac{uccsd} and the \ac{kupccgsd} ansatz.

The \ac{uccsd} ansatz truncates the \ac{ucc} ansatz by only keeping the singles and doubles
term in \cref{eq:uccwave},
\begin{equation}
  \label{eq:uccsd-ansatz}
  U_{\textrm{ansatz}} = \exp\left(\hat{T_{1}}-\hat{T_{1}}^{\!\!\dagger} + \hat{T_{2}}-\hat{T_{2}}^{\!\!\dagger}\right) \ .
\end{equation}
In order to physically realize this many-body unitary operator, we have to break it into
one and two body operator. Since the excitation operators do not commute,
$[\hatT_{i},\hatT_{j}] \neq 0$, we apply the \ac{ts}
decomposition~\cite{poulin_2014_trotterstep}
\newcommand{\tidag}{\hat{T_i}^{\!\dagger}}
\begin{equation}
  \label{eq:trotter}
  e^{\sum_{i} \hatT_{i}- \tidag} = \left( e^{\sum_{i} (\hatT_{i}- \tidag)/n}\right)^{n}
   \approx (\prod_{i}e^{(\hatT_{i}- \tidag)/n})^{n}
\end{equation}
to simplify the unitary. The first order \ac{ts} expansion, $n=1$, suffices for the
purpose. Since the individual terms of $\hatT_{1}$ and $\hatT_{2}$ commute with each
other, \cref{eq:trotter} can be further simplified as
\begin{equation}
  \label{eq:uccsd-circuit}
  U_{\textrm{ansatz}} = \prod_{i,a}e^{t^{a}_{i}a_{a}^{\dagger}a_{i}-\text{h.c.}}\prod_{ij,ab}e^{t^{ab}_{ij}a^{\dagger}_{a}a^{\dagger}_{b}a_{j}a_{i}-\text{h.c.}}
\end{equation}
where the indices $i$ and $j$ run over the $\eta$ occupied orbitals, and indices $a$ and
$b$ run over the $M-\eta$ virtual orbitals. Each of these unitaries can be applied one after
the other on a quantum computer. To prepare this ansatz on a quantum computer, we first
translate the fermionic operators $a$ and $a^{\dagger}$ to their corresponding Pauli operators
that act in the qubit basis. We provide details of such transformation in Methods. From
its form in \cref{eq:uccsd-circuit}, it becomes evident that the \ac{uccsd} ansatz of
molecule with $\eta$ active electrons and $M$ active orbitals has
$O((M-\eta)^{2}\eta^{2})$ parameters. In its simplest
implementation~\cite{whitfield_2011_hamqpe}, the common \ac{jw} encoding adds another
$O(M)$ overhead (See Methods, \ref{method:standard-circuit}). Thus, the number of gates
required to implement \ac{uccsd} ansatz via \ac{vqe} scales as
$O(M(M-\eta)^{2}\eta^{2})$. However, simple optimization via rearrangement of terms can remove
the \ac{jw} overhead which leads to optimal scaling of gates in \ac{uccsd}
ansatz~\cite{hastings_2015_improvingquantum}. In \cref{fig:1}, we show the number of gates
required to implement the \ac{uccsd} ansatz with and without the \ac{jw} overhead.

This scaling of the \ac{vqe} algorithm can be improved by considering other ansatz instead
of the \ac{uccsd} ansatz. Low depth or linear scaling ansatz can significantly reduce the
resource requirement while maintaining a similar level of
accuracy~\cite{kivlichan_2018_quantumsimulation,lee_2019_kUpCCGSD}. As an example, we
consider the \ac{kupccgsd} ansatz introduced in Ref.~\cite{lee_2019_kUpCCGSD} The
\ac{kupccgsd} ansatz is implemented by the unitary
\begin{equation}
  \label{eq:kupccgsd-ansatz}
  U_{\textrm{ansatz}} = \prod_{i=1}^{k} \exp\left( \hat{T_{1}}^{(i)} + \hat{T_{2}}^{(i)} - \text{h.c.} \right)
\end{equation}
where
\begin{align}
  \label{eq:generalized-sd}
  \begin{split}
    \hatT_{1}^{(i)} &= \sum_{p,q=1}^{M} {t_{q}^{p}}^{(i)} a^{\dagger}_{p} a_{q} \\
    \hatT_{2}^{(i)} &= \sum_{p,q=1}^{M/2} {t_{q_{\alpha}q_{\beta}}^{p_{\alpha}p_{\beta}}}^{(i)} a^{\dagger}_{p_{\alpha}}a^{\dagger}_{p_{\beta}} a_{q_{\beta}}a_{q_{\alpha}}
  \end{split}
\end{align}
are $k$ copies of the generalized paired single and doubles excitation operators. In this
ansatz, the orbitals are no longer separated into occupied and virtual orbitals, single
electron excitations are allowed between any pair of orbitals and double excitations are
allowed from one spatial orbital to other. The pairing of electrons for double excitations
greatly reduces the number of parameters in the \ac{kupccgsd} ansatz which scales as
$O(kM^{2})$. In \cref{fig:1}, we show the number of two-qubit gates required to implement
the unitary with $k=1$. As expected, the \ac{kupccgsd} ansatz require an order of
magnitude fewer gates than the \ac{uccsd} ansatz. The parameter $k$ needs to be tuned
experimentally. Initial work shows that $k$ does not scales as fast as
$M$~\cite{lee_2019_kUpCCGSD}, but more work will be required to find an optimal $k$ value
for dipeptide simulation.

\subsection*{Circuit Depth}
\label{sec:circuit-depth}
\begin{table}[t]
\centering
\kupccgsdDepth{}
\caption{\textbf{Approximate circuit depth to implement \ac{kupccgsd} ansatz.} We estimate
  the approximate circuit depth by scaling gate count by a factor of $O(M)$ where $M$ is
  the number of spin orbitals used to simulate each dipeptide. We show the $k=1$ case.}
\label{tab:3}
\end{table}

The circuit depth of these quantum circuits will depend on the number of gates that can be
applied simultaneously during preparation of the ansatz. In simplest form, the gates that
act on different qubits of a wave function can be applied simultaneously. For example, we
can prepare the \ac{kupccgsd} ansatz via the circuit denoted by \cref{eq:uccsd-circuit} by
simultaneously applying terms with coefficients $t_{qq}^{pp}$ and $t_{q'q'}^{p'p'}$ as
long as they act on different qubits. If the spin orbitals of the \ac{kupccgsd} ansatz are
labeled such that consecutive qubits represent the $\alpha$ and $\beta$ spin orbitals, then the
terms where $p<p'<q'<q$ can be applied simultaneously. In this case, the term with
coefficient $(p',q')$ is nested inside the term with coefficient $(p,q)$ and hence can be
applied simultaneously~\cite{hastings_2015_improvingquantum}. In general, the circuit
depth is a factor of $O(M)$ smaller than the number of gates used to implement the
ansatz~\cite{lee_2019_kUpCCGSD} (See Methods, \ref{method:nesting-terms}). In
\cref{tab:3}, we show the approximate circuit depth of the \ac{kupccgsd} ansatz
implemented for several dipeptide simulations. These circuits have a circuit depth of
roughly a few thousands at $k=1$, but practical implementation of these ansatz might
require a higher value of $k$~\cite{lee_2019_kUpCCGSD}.
%

%%%%%%%%%%%%%%%%%%%%%%%%%%%%%%%%%%%%%%%%%%%%%%%%%%%%%%%%%%%%%%%%%%%%%%%%%%%%%%%%%
\section{Discussion}
\label{sec:discuss}
%%%%%%%%%%%%%%%%%%%%%%%%%%%%%%%%%%%%%%%%%%%%%%%%%%%%%%%%%%%%%%%%%%%%%%%%%%%%%%%%%

Quantum computing holds great promise of improving accuracy and the scale of numerical
simulations used in chemistry and other sciences. In particular, quantum computing can
enable high quality \abinitio simulations at a larger scale than possible via current
classical computational techniques. An accurate computation via \abinitio methods remains
the only viable tool for quantitative analysis of a system where quantum effects dominate,
such as those with few atoms, where bond breaking/formation takes place, etc. At larger
scale, \ac{md} simulations can provide accurate predictions for chemical reactions,
provided one starts with a high quality force field. We have discussed the use of quantum
computing and related \abinitio simulation capabilities to tie these two approaches
together, where results from quantum computing simulations can guide the development of
better force fields. As an illustrative example, we have provided quantum resource
estimates for performing dipeptide simulation, a task of direct importance for optimizing
protein force fields. This task requires a few hundred qubits and a circuit depth of few
thousands. Current generation of quantum computers, with around 50 qubits and a possible
circuit depth of a thousand~\cite{google_2019_supremacy} are only an order of magnitude
away from this requirement. Thus, these computations with active space reduction are
feasible to attain on the \ac{nisq} era quantum computers.

Our work comes with some important caveats. We have assumed that we have access to
a quantum computer with all to all qubit connectivity and that these qubits can be
controlled via single qubit rotations and CNOT gates. This allows us to assume that we
can assign a spin orbital to an arbitrary qubit. However, regular quantum hardware
has a connectivity graph with finite degree and additional qubits and gates will be
required to overcome this limited connectivity. Ancillary qubits might also be required to
perform required quantum gates, further increasing the qubit count. Thus, our work should
be taken as a lower bound towards the required quantum resources. Nevertheless, the
scaling of the circuit depth and qubit count of an optimal ansatz will remain linear in the
size of molecules.

This linear scaling allows us to consider further improvement for protein force field
parameterization beyond dipeptide simulations, such as tripeptides
simulation~\cite{anishetty_2002_tripeptideanalysis, culka_2019_initioprotein}. Simulations
of larger peptide structure might reveal terms in a protein force field which depend on
higher order interaction of peptides. \Abinitio data can be used to improve other forms of
force fields, such as \ac{aiff}~\cite{xu_2018_aiff} and
\ac{nnp}~\cite{smith_2017_ani1,chmiela_2018_mlff} force field. These ideas serve as future
extensions to this work.

%%%%%%%%%%%%%%%%%%%%%%%%%%%%%%%%%%%%%%%%%%%%%%%%%%%%%%%%%%%%%%%%%%%%%%%%%%%%%%%%%
%%%%%                        METHODS                                     %%%%%%%%
%%%%%%%%%%%%%%%%%%%%%%%%%%%%%%%%%%%%%%%%%%%%%%%%%%%%%%%%%%%%%%%%%%%%%%%%%%%%%%%%%
\section*{Methods}
\label{sec:methods}

\subsection*{Jordan Wigner transform}
\customlabel{method:jw-transform}{Jordan Wigner transform}

The electronic Hamiltonian, \cref{eq:2quantHam} is written in terms of indistinguishable
fermions, while the qubits of a quantum computer are distinct registers with no specific
spin. We require a transform that will allow us to represent electronic system on qubits.
This task is called encoding, where the fermionic operators are rewritten as a string of
operators on qubits. In the \ac{jw} encoding, each spin orbital is represented by a
qubit. If the spin orbital is occupied, the qubit is set to state $\ket{1}$ and it is set
to $\ket{0}$ otherwise. If the size of our basis set is $M$, then we require $M$ qubits to
encode the wave function. The vacuum state $\ket{\Psi} = \ket{0}$ is represented as
$\ket{\Psi}=\ket{0}^{\otimes M}$ in qubit representation, while state $\ket{ij}$ with one electron
each in orbital $i$ and $j$ is written as
\begin{equation}
  \label{eq:jw-state}
  \ket{\Psi} = \ket{ij} \equiv \ket{000\ldots0\underset{i}{1}0\ldots00\underset{j}{1}00\ldots0}
\end{equation}
where the subscripts denote the index of the orbital in qubit notation. Next, we have to
specify the creation $a_{i}^{\dagger}$ and annihilation operators $a_{i}$ in terms of single
qubit gates. The creation operator should converts qubit $\ket{0}$ to $\ket{1}$,
vice--versa for the annihilation operator. They are given by
\begin{align}
  \label{eq:jw-transform}
  a_{i} &= \sigma^{z}_{1}\otimes \sigma_{2}^{z} \otimes \ldots \otimes \sigma^{z}_{i-1} \otimes {(\ketbra{0}{1})}_{i} \\
  a_{i}^{\dagger} &= \sigma^{z}_{1}\otimes \sigma_{2}^{z} \otimes \ldots \otimes \sigma^{z}_{i-1} \otimes {(\ketbra{1}{0})}_{i}
\end{align}
The product of $\sigma^{z}$ operator maintains the proper antisymmetry of the fermionic
operators, and is referred to as the \ac{jw} string. The basis vectors in the \ac{jw}
encoding are simple representation of electron occupation, but fermionic operators become
many-qubit gates.

\subsection*{Standard circuit}
\customlabel{method:standard-circuit}{Standard circuit}

\begin{figure}
  \centerline{\input{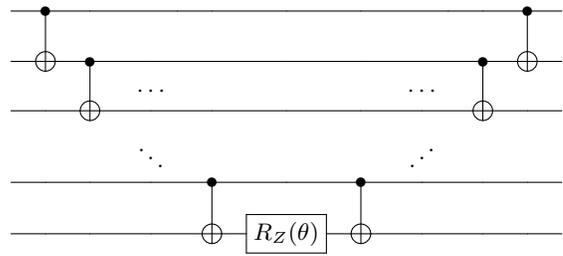}}
  \caption{Quantum circuit that implements the unitary $\exp(\ii \theta Z_1 Z_2 \ldots Z_n)$}
  \label{fig:2}
\end{figure}

Here, we shall summarize the quantum circuit used by Ref.~\cite{whitfield_2011_hamqpe} to
implement fermionic operations with \ac{jw} strings. After \ac{jw} encoding, various
\ac{vqe} ansatz implement exponentiated strings of Pauli operators, such as a string of
$Z$ operators on $n$ qubits:
\begin{equation}
  \label{eq:z-string-exp}
  U = \exp(\ii \theta Z_{1}Z_{2}\ldots Z_{n}) \ .
\end{equation}
In \cref{fig:2}, we show the circuit that can implement this unitary only via a basic single
qubit gate and CNOT gates. Pauli strings with other Pauli operators (such as $X$ or $Y$)
can be handled by using the identities
\begin{align}
  \label{eq:xy-string}
  \begin{split}
    \e^{\ii X} &= H \e^{\ii Z} H \\
    \e^{\ii Y} &= Y \e^{\ii Z} Y
  \end{split}
\end{align}
where $H$ and $Y$ are the Hadamard and $Y$ single qubit gate respectively (See
\cref{sec:common-gates}). Thus, we can implement all unitaries of exponentiated
Pauli strings by combining single qubit gates and the circuit shown in \cref{fig:2}.

\subsection*{Nesting terms}
\customlabel{method:nesting-terms}{Nesting terms}

In this section, we want to compute the scaling factor between the circuit count and the
circuit depth of the \ac{kupccgsd} ansatz. The total number of terms in operator
$\hat{T}_{2}$ of \cref{eq:generalized-sd} is $\frac{M}{2}(\frac{M}{2}-1)$ and hence the
gate count of the circuit scales as $O(M^{2})$. Two terms $(p,q)$ and $(p',q')$ of this
operator can be applied simultaneously if they are nested inside each other such that
$p>p'>q'>q$~\cite{hastings_2015_improvingquantum}.  Let $S(p,q)$ be the set of consecutive
terms separated by one spatial orbital that can be nested inside the term $(p,q)$, that
is,
\begin{align}
  \label{eq:nested-set}
  \begin{split}
    S(p,q) &= \{ (p,q), (p-1,q+1), (p-2,q+2), \ldots \} \\
           &\equiv \{ (l,m)\ |\ l + m = p + q,\ l>m \} \ .
  \end{split}
\end{align}
The number of such distinct disjoint sets is the number of possible values of
$p+q$ where $p, q \in \{1, 2, 3, \ldots, M/2\}$ and $M$ is the number of distinct spin
orbitals. The distinct values of $p+q$ ranges between $3$ to $M-1$, and the number
of such distinct sets is $M-3$. So, the circuit depth of the \ac{kupccgsd} ansatz scales
as $O(M)$, which is an $O(M)$ improvement over its gate count which scales as $O(M^{2})$.

\noindent\textbf{Acknowledgement.} We thanks Shu Ching Ou, Micheal Gilson and Evgeny Epifanovsky for useful discussion.

\newcommand{\Eprint}[2]{\href{#1}{\textcolor{RedOrange}{#2}}}%
% Bibliography

%%%%%%%%%%%%%%%%%%%%%%%%%%%%%%%%%%%%%%%%%%%%%%%%%%%%%%%%%%%%%%%%%%%%%%%%%%%%%%%%%
%%                   SUPPLEMENTARY INFORMATION                               %%%%
%%%%%%%%%%%%%%%%%%%%%%%%%%%%%%%%%%%%%%%%%%%%%%%%%%%%%%%%%%%%%%%%%%%%%%%%%%%%%%%%%
\appendix
\acresetall % Reset all acronyms
\renewcommand\thesubsection{\arabic{subsection}}
\makeatletter
\renewcommand\p@subsection{\thesection}
\makeatother
\titleformat{\section}
  {\sffamily\bfseries}
  {\thesection.}
  {0.5em}
  {#1}
  []

\section{Hartree-Fock theory}
\label{sec:hf}

In principle, all the properties of a molecular system can be obtained from its
wave function which itself can be obtained by solving the Schr\"{o}dinger equation. Since
electrons are much lighter than atomic nuclei, the Born-Oppenheimer approximation
neglects the motion of the nuclei. After applying the approximation, the electronic
Hamiltonian is given by the sum of electronic kinetic energy and the electron-electron and
nuclei-electron Coulomb interaction energy;
\begin{equation}
  \label{eq:1}
  H(\{\vec{r}_{i}\}) =-\frac{1}{2} \sum_{i=1}^{n} \nabla_{i}^{2}
  + \sum_{i>j}^{n} \frac{1}{\left|\vec{r}_{i}-\vec{r}_{j}\right|}
  - \sum_{i=1}^{n}\sum_{\alpha=1}^{N_{\rm n}} \frac{Z_{\alpha}}{\left|\vec{r}_{i}-\vec{r}_{\alpha}\right|}
\end{equation}
where $n$/$N_{\rm n}$ is the number of electrons/nuclei in the molecule,
$\vec{r}_{i}/\vec{r}_{\alpha}$ are the electronic/nuclear coordinates and $Z_{\alpha}$ are the
nuclear charges. The eigenfunctions to this Hamiltonian are the molecular electronic
wave-function of the ground state and various excited states, and the eigenvalues are the
corresponding energies. The ground state energy as a function of electronic coordinates,
$E(\{\vec{r}_{i}\})$, is called the \ac{pes} of the molecule. In
general, one cannot find an analytical solution to a many body Schr\"{o}dinger
equation. However, approximate computational methods can yield results which are with-in
chemical accuracy.

Let $\{\phi_{j}(\vec{r})\}$ be a complete basis set such that the wave function of the $i^{\rm
th}$ electron can be written as
\begin{equation}
  \label{eq:2}
  \psi_{i}(\vec{r}) = \sum_{j}c_{ij}\phi_{j}(\vec{r})
\end{equation}
where $\mytensor{C} = [c_{ij}]$ are appropriate coefficients. The possible many-electron
states can be constructed from these single electron wave functions. Since electrons are
fermions, the many-electron state must be antisymmetric in any two electron
coordinates. A many-electron state where the electrons occupy states
$\{\psi_{i_{1}},\psi_{i_{2}},\ldots,\psi_{i_{n}}\}$ can be written as a Slater's determinant
\begin{equation}
  \label{eq:3}
  \Psi(\vec{R},\mytensor{C}) =
  \begin{vmatrix}
    \psi_{i_{1}}(\vec{r}_{1}) & \psi_{i_{1}}(\vec{r}_{2}) & \ldots & \psi_{i_{1}}(\vec{r}_{n}) \\
    \psi_{i_{2}}(\vec{r}_{1}) & \psi_{i_{2}}(\vec{r}_{2}) & \ldots & \psi_{i_{2}}(\vec{r}_{n}) \\
    \vdots & & & \\
    \psi_{i_{n}}(\vec{r}_{1}) & \psi_{i_{n}}(\vec{r}_{2}) & \ldots & \psi_{i_{n}}(\vec{r}_{n})
  \end{vmatrix}
\end{equation}
where $\vec{R} = \{\vec{r}_{1}, \vec{r_{2}}, \ldots ,\vec{r_{n}}\}$ represent the coordinates
of the electrons. Note that this state depends on the coefficient matrix
$\mytensor{C}$. In general, the actual electronic wave function can be written as an
appropriately weighted sum of these determinants. The \ac{hf} approximation assumes that
the wave function consists of only one such determinant, and then optimizes the parameters
$\mytensor{C}$ of the determinant by applying the variational principle. We want to find
the optimal matrix $\mytensor{C}$ that minimizes the expectation energy of the state
$\Psi(\vec{R},\mytensor{C})$ while ensuring that the \ac{mo} $\psi_{i}$ are
appropriately normalized. We can do this by introducing a set of Lagrange multiplier
$\epsilon_{i}$ such that
\begin{equation}
  \label{eq:4}
  \delta \left[ \int \Psi^{*} H \Psi\ \mathrm{d}\vec{R} - \sum_{i}\epsilon_{i}\int |\psi_{i}|^{2} \mathrm{d}\vec{r}_{i} \right] = 0
\end{equation}
This leads to a set of single-electron coupled equations, collectively called the
Hartree-Fock equations:
\begin{widetext}
  \begin{equation}
    \label{eq:hf}
    \left[ -\frac{1}{2}\nabla^{2}_{i} - \sum_{i\alpha} \frac{Z_{\alpha}}{|\vec{r}_{i}-\vec{r}_{\alpha}|} \right]   \psi_{i}(\vec{r}_{i})
    + \sum_{j\neq i}\left[ \int \mathrm{d}r_{j} \frac{|\psi_{j}(\vec{r}_{j})|^{2}}{|\vec{r}_{i}-\vec{r}_{j}|} \right] \psi_{i}(\vec{r}_{i})
    - \sum_{j\neq i}\left[ \int \mathrm{d}r_{j} \frac{\psi_{j}(\vec{r}_{j}) \psi_{i}(\vec{r}_{i})}{|\vec{r}_{i}-\vec{r}_{j}|} \right] \psi_{i}(\vec{r}_{i})
    = \epsilon_{i} \psi_{i}(\vec{r}_{i})
  \end{equation}
\end{widetext}
The first term on the left hand side is the sum of electron kinetic energy and
electron-nuclei Coulomb interaction, the second term is the electronic interaction of
electron $i$ with the mean field electric force of all other electrons and the third term
is the exchange energy term. Electrons of the same spin avoid each other due to Pauli's
exclusion principle and experience smaller Coulomb repulsion. This gives rise to the
exchange energy term.

The \ac{hf} equations~\eqref{eq:hf} are highly non-linear as $\psi_{i}$ features on both
sides of equality. The \ac{hf} equations are mean-field single electron equations and
require the wave functions of all other electrons to write down the Coulomb and exchange
energy. The \ac{scf} method is often used to solve these equations. The \ac{scf} method
starts with a reasonable guess of the single electron wave function, and solves the \ac{hf}
equations assuming the guess wave function for all the other electrons. This solution is
then used as the wave function of other electrons, and the HF equations are solved again,
yielding a hopefully better wave function. At each step, we keep track of the energy of the
solution, $E=\sum_{i}\epsilon_{i}$. The \ac{scf} method stops when the energy converges.

%%%%%%%%%%%%%%%%%%%%%%%%%%%%%%%%%%%%%%%%%%%%%%%%%%%%%%%%%%%%%%%%%%%%%%%%%%%%%%%%%
\section{Second Quantization}
\label{sec:2quant}

The \ac{hf} wave function is a mean-field approximation of the ground state many-body
electronic wave function. Since it is constructed from a single determinant, it ignores
static and dynamic electron-electron interaction. Post \ac{hf} methods try to recover
this correlation by considering additional electronic configurations. These methods
becomes considerably easier to analyze in the second quantization formulation of the
Hamiltonian problem. In this section, we describe the method of obtaining second quantized
Hamiltonian from variationally optimized \ac{hf} functions.

We start with the variationally optimized \ac{hf} molecule orbitals, $\{\psi_{i}\}$, and
represent the Slater determinant~\eqref{eq:3} in the Dirac notation as
\begin{equation}
  \label{eq:6}
  \ket{\Psi} \equiv \ket{i_{1}i_{2}\ldots i_{n}}
\end{equation}
This is a $n$-electron state, where the electrons occupy the orbitals
$\{i_{1},i_{2},\ldots,i_{n}\}$ while the other molecular orbitals are empty. In this notation
$\ket{\phantom{i}}$ denotes the state of no electron. The excitation operator
$a^{\dagger}_{i}$ creates an electron in one of the orbital:
\begin{equation}
  \label{eq:7}
  \ket{i\, j_{1} j_{2} \ldots } = a^{\dagger}_{i}\ket{j_{1} j_{2} \ldots} \ .
\end{equation}
If state $i$ is already occupied, $a_{i}^{\dagger}\ket{\Psi}=0$. Any state $\ket{\Psi}$ can be
uniquely specified by a string of creation operators,
\begin{equation}
  \label{eq:8}
  \ket{\Psi} = a^{\dagger}_{i_{1}}a^{\dagger}_{i_{2}}\ldots a^{\dagger}_{i_{n}}\ket{0}
\end{equation}
Similarly, the annihilation operator $a_{i}$ destroys the electron in one of the
orbital:
\begin{equation}
  \label{eq:9}
  \ket{j_{1} j_{2} \ldots } = a_{i}\ket{i\ j_{1} j_{2} \ldots } \ .
\end{equation}
If the state $i$ is already unoccupied, then $a_{i}\ket{\Psi}=0$. These definitions, along
with the completely antisymmetric nature of the wave function $\ket{\Psi}$ imply that these
operators obey the canonical anticommutation relationship,
\begin{align}
  \label{eq:10}
  \{a^{\dagger}_{i},a_{j}\} &= a^{\dagger}_{i}a_{j} + a_{j}a^{\dagger}_{i} = 0 \\
  \{a_{i},a_{j}\} &= \{a^{\dagger}_{i},a^{\dagger}_{ij}\} = 0
\end{align}
We can now rewrite the Hamiltonian in its second quantized form:
\begin{equation}
  \label{eq:11}
  \hat{H} = \sum_{ij} h_{ij}a^{\dagger}_{i}a_{j} + \sum_{ijkl} V_{ijkl} a^{\dagger}_{i}a^{\dagger}_{j}a_{l}a_{k}
\end{equation}
where
\begin{align}
  \label{eq:12}
  h_{ij} &= \int \psi^{*}_{i}(\vec{r}) \left[ -\frac{1}{2} \nabla^{2} - \sum_{\alpha} \frac{Z_{\alpha}}{|
           \vec{r}-\vec{r}_{\alpha}|}\right] \psi_{j}(\vec{r})\, \mathrm{d}\vec{r} \ , \\
  \label{eq:13}
  V_{ijkl} &= \int \psi^{*}_{i}(\vec{r}_{1})\psi^{*}_{j}(\vec{r}_{2})
             \frac{1}{|\vec{r}_{1}-\vec{r}_{2}|} \psi_{k}(\vec{r}_{2})\psi_{l}(\vec{r}_{1}) \,
             \mathrm{d}\vec{r}_{1}\mathrm{d}\vec{r}_{2} \ .
\end{align}

In the limit where the \ac{mo}s $\{\psi_{i}\}$ form a complete basis set, the second
quantized representation is exact. In computational chemistry applications, frequently the
basis set is limited to $M$ functions, in which case second quantized
Hamiltonian~\eqref{eq:11} is only an approximate representation the electronic
Hamiltonian~\eqref{eq:1}. The Hamiltonian contains $\bigO({M^{4}})$ terms, and becomes
increasing hard to solve via exact diagonalization.

%%%%%%%%%%%%%%%%%%%%%%%%%%%%%%%%%%%%%%%%%%%%%%%%%%%%%%%%%%%%%%%%%%%%%%%%%%%%%%%%%
\section{Post-Hartree Fock methods}
\label{sec:posthf}

The \ac{hf} solution serves as a good starting point for post-\ac{hf} methods, which
iterate over the \ac{hf} solution and show better agreement with experimental results. In
this section, we discuss two such post-Hartree Fock methods.

\subsection*{Configuration interaction method}
\label{sec:ci}

A general solution of the full Hamiltonian~\eqref{eq:1} can be constructed by taking an
appropriate weighted sum of all possible determinants. This is known as the \ac{fci}
method where the wave function is given by
\begin{equation}
  \label{eq:14}
  \ket{\Psi_{\rm FCI}} = \sum c_{i} \ket{\Psi_{i}}
\end{equation}
The sum is done over all possible determinants of form given by \cref{eq:3}. If these
determinants are constructed out of $M$ molecular orbitals with $N$ electrons, the number
of possible determinants is ${M \choose N} = \frac{M!}{(M-N)!N!}$ which scales
exponentially in $M$. Thus, an \ac{fci} calculation is intractable except for very small
molecules.

The \ac{fci} calculation can be simplified by using additional symmetries of the
molecule. Hamiltonian~\eqref{eq:1} commutes with the spin operators $\hat{S}_{z}$ and
$\hat{S}^{2}$ and also commutes with the z-component of the total angular momentum
$\hat{L}_{z}$, that is, the Hamiltonian preserves the total intrinsic spin of
electrons and the z-component of the total angular momentum. Thus, the valid
determinants in \cref{eq:14} should be of same spin. Such a combination of the determinants
is called a \ac{csf}. Using a \ac{csf} can greatly reduce the number of determinants
required to construct an \ac{fci} wave function. Nevertheless, the number of determinants in
\ac{csf} wave function also grow exponentially in the size of basis set, and \ac{fci} with
such methods becomes intractable for large molecules.

In second quantization formulation, the \ac{fci} wave function can be constructed
systematically from the \ac{hf} state. Let $\ket{\Psi_{\rm HF}}$ be the \ac{hf}
determinant~\eqref{eq:3} constructed from molecular orbitals found via the \ac{scf}
method. The \ac{fci} wave function is constructed via a series of excitation
operators:
\begin{equation}
  \label{eq:15}
  \ket{\Psi_{\rm FCI}} = \left( \sum_{ia}c^{a}_{i}a^{\dagger}_{a}a_{i} +
    \sum_{ij,ab}c^{ab}_{ij}a^{\dagger}_{a}a^{\dagger}_{b}a_{i}a_{j} + \cdots \right) \ket{\Psi_{\rm HF}}
\end{equation}
where the indices $\{i,j,\cdots\}$ and $\{a,b,\cdots\}$ run over the occupied and unoccupied
orbitals respectively in the $\ket{\Psi_{\rm HF}}$ wave function. The different determinants
in the \ac{fci} expansion are classified as singles (S), doubles (D), triples (T), etc
depending on their level of excitation from the \ac{hf} wave function. The \ac{fci}
wave function can be systematically approximated via the \ac{ci} method by inclusion of
different levels of excitations, for example, \ac{cisd} keeps singles and doubles, CISDTQ
also adds triple and quadruple excitations, etc.

\subsection*{Coupled cluster method}
\label{sec:cc}

The \ac{ci} formulation is exact under the complete basis set limit (i.e.
$M\rightarrow\infty$), but is not size extensive and converges very slowly to the full wave
function. These shortcomings can be overcome by the \ac{cc} method. The \ac{cc}
wave function is given by
\begin{equation}
  \label{eq:16}
  \ket{\Psi_{\rm CC}} = \exp\left(\sum_{i,a}t^{a}_{i} a^{\dagger}_{a}a_{i} +
    \sum_{ij,ab} t^{ab}_{ij} a^{\dagger}_{a}a^{\dagger}_{b}a_{j}a_{i} +
    \cdots \right) \ket{\Psi_{\rm HF}}
\end{equation}
where the indices $i,j,\ldots$ run over the occupied levels and $a,b,\ldots$ run over the unoccupied
level. If all excitation levels are included, then the \ac{cc} and \ac{fci} expansions
describe the same wave function. For example, by expanding and comparing the terms in
\cref{eq:15,eq:16}, we find that
\begin{align}
  \label{eq:17}
  c^{a}_{i} &= t^{a}_{i} \ , \\
  c^{ab}_{ij} &= t^{ab}_{ij} + \frac{1}{2}(t^{a}_{i}t^{b}_{j} + t^{a}_{j}t^{b}_{i}) \ ,
\end{align}
etc. The \ac{cc} wave function is size extensive and usually converges faster than the
\ac{ci} wave function~\cite{bartlett_2007_cctheory}.

We want to solve the \ac{cc} Schr\"{o}dinger equation $\hat{H}\ket{\Psi_{\rm CC}} =
E\ket{\Psi_{\rm CC}}$. We define the coupled cluster operator
\begin{align}
  \label{eq:18}
  \hat{T} &= \hat{T_{1}} + \hat{T_{2}} + \ldots \\
          &= \sum_{i,a}t^{a}_{i} a^{\dagger}_{a}a_{i} + \sum_{ij,ab} t^{ab}_{ij} a^{\dagger}_{a}a^{\dagger}_{b}a_{j}a_{i} + \cdots
\end{align}
such that
\begin{equation}
  \label{eq:19}
  \ket{\Psi_{\rm CC}} = \e^{\hat{T}} \ket{\Phi_{0}}
\end{equation}
where $\ket{\Phi_{0}}$ is the starting wave function, usually a \ac{hf} wave function or
a \ac{csf}. We want to solve for the energy $E$ and the amplitudes
$\{t^{a}_{i},t^{ab}_{ij},\ldots\}$. Noting that the determinants
$\ket{\Phi^{ab\ldots}_{ij\ldots}} = a^{\dagger}_{a}a^{\dagger}_{b}\ldots a_{j}a_{i}\ket{\Phi_{0}}$ form an orthonormal set,
we can write down the coupled cluster equations:
\begin{align}
  \label{eq:20}
  \braket{\Phi_{0}|\e^{-\hat{T}}\hat{H}e^{\hat{T}}|\Phi_{0} } &= E \\
  \braket{\Phi^{ab\ldots}_{ij\ldots}|\e^{-\hat{T}}\hat{H}e^{\hat{T}}|\Phi_{0}} &= 0
\end{align}
The operator $\e^{-\hat{T}}\hat{H}e^{\hat{T}}$ can be simplified via the \ac{bch}
expansion~\cite{eichler_1968_proofbch} and by noting that this series terminates because
the Hamiltonian $\hat{H}$ as written in \cref{eq:11} only contains 2-excitation
operators. Thus,
\begin{align}
  \begin{split}
    \label{eq:21}
    \e^{-\hat{T}}\hat{H}e^{\hat{T}} &= \hat{H} + [\hat{H}, \hat{T}] + \frac{1}{2!}[\hat{H},[\hat{H}, \hat{T}]] + \frac{1}{3!} [\hat{H},[\hat{H},[\hat{H}, \hat{T}]]] \\
    &+ \frac{1}{4!} [\hat{H}, [\hat{H}, [\hat{H},[\hat{H}, \hat{T}]]]] \ .
  \end{split}
\end{align}

The above technique describes a projective method of solving the \ac{cc} equations. The
projective \ac{cc} equations are convenient to solve via numerical methods. The energy
formula in \cref{eq:20} does not conform to variational condition as the operator
$\e^{-\hat{T}}\hat{H}e^{\hat{T}}$ is not Hermitian. Thus, energy found via solving
\cref{eq:20} with a truncated operator $\hat{T}$ might not be an upper bound to the true
coupled cluster energy. We can construct a variational solution by starting from
\cref{eq:19,eq:18} and using the Hermitian conjugate of $\e^{\hat{T}}$. This yields a
variational form such that for truncated operator $\hat{\tau}=\sum_{i}^{n}T_{i}$,
\begin{equation}
  \label{eq:22}
  E = \frac{\braket{\Phi_{0}|(\e^{\hat{T}})^{\dagger}\hat{H}e^{\hat{T}}|\Phi_{0}}}
  {\braket{\Phi_{0}|(\e^{\hat{T}})^{\dagger}\e^{\hat{T}}|\Phi_{0}}}
  \leq \tilde{E} =
  \frac{\braket{\Phi_{0}|(\e^{\hat{\tau}})^{\dagger}\hat{H}e^{\hat{\tau}}|\Phi_{0}}}
  {\braket{\Phi_{0}|(\e^{\hat{\tau}})^{\dagger}\e^{\hat{\tau}}|\Phi_{0}}} \ .
\end{equation}
The operator $(\e^{\hat{\tau}})^{\dagger}\hat{H}e^{\hat{\tau}}$ does not have a known finite length
expansion which makes finding solutions of \cref{eq:22} a considerably harder
computational task when compared to the projective methods. Nevertheless, it can be
approximated by expanding $e^{\hat{\tau}}$ via a Taylor series and then truncating it. Such
truncation is arbitrary and leads to additional errors.

A similar variational method called the \ac{ucc} method tries a different approach. The
cluster operator $\hat{T}$ is replaced by an anti-Hermitian operator
$\hat{T}-\hat{T}^{\dagger}$ such that $\hat{U} = \e^{\hat{T}-\hat{T}^{\dagger}}$ is a unitary
operator. The resultant operator $\hat{U}^{\dagger}\hat{H}\hat{U}$ in
\begin{equation}
  \label{eq:23}
  E = \braket{\Phi_{0}|\hat{U}^{\dagger}\hat{H}\hat{U}|\Phi_{0}}
\end{equation}
can be thought as a rotation of basis such that
$\hat{H^{'}} = \hat{U}^{\dagger}\hat{H}\hat{U}$ is a Hamiltonian that has the same eigenvalues
as $\hat{H}$. A systematic series expansion of $\hat{H'}$ can be obtained where we
truncate the series by keeping all terms to a particular order of perturbation
theory~\cite{hoffmann_1988_UCC}. Alternatively, the energy
$E=\braket{\Phi'|\hat{H}|\Phi'}$ can be thought as the expectation value of the Hamiltonian with
respect to the wave function
\begin{equation}
  \label{eq:24}
  \ket{\Phi'} = \hat{U} \ket{\Phi_{0}}
\end{equation}
This approach is suitable to quantum computing where a universal quantum computer can
efficiently prepare the state $\ket{\Phi'}$ by applying the unitary $\hat{U}$ on an initial
state $\ket{\Phi_{0}}$.

%%%%%%%%%%%%%%%%%%%%%%%%%%%%%%%%%%%%%%%%%%%%%%%%%%%%%%%%%%%%%%%%%%%%%%%%%%%%%%%%%
\section{\Acl*{vqe}}
\label{sec:vqe}

The \ac{vqe} method repeatedly prepares a variational quantum state $\ket{\Psi(\vec{t})}$
and optimizes over the unknown parameters $\vec{t}$ to estimate the ground state energy. A
quantum computer can be used to efficiently prepare the state $\ket{\Psi(\vec{t})}$ and the
energy of this state $E(\vec{t})=\braket{\Psi(\vec{t})|H|\Psi(\vec{t})}$ can be estimated on a
classical computer via repeated measurements of the quantum state. A classical optimizer
optimizes the parameters $\vec{t}$ to minimize energy $E(\vec{t})$. Thus, \ac{vqe} is a
hybrid classical-quantum algorithm.

Different encoding methods convert the fermionic Hamiltonian to a qubit-based Hamiltonian
which is a weighted sum of strings of Pauli operators,
\begin{equation}
  \label{eq:39}
  H = \sum_{i} H_{i}\, \prod_{\bm{\sigma}_{i}} \sigma^{\alpha}_{j}
\end{equation}
where $\alpha\in\{x,y,z\}$ specifies one of the Pauli operators and
$\sigma^{\alpha}_{j} \in \bm{\sigma}_{i}$ where $\bm{\sigma}_{i}$ is a set of Pauli operators describing the
Pauli string of $i^{\rm th}$ term. The parameters $H_{i}$ depend on the parameters
$h_{ij}$ and $V_{ijkl}$ computed via \cref{eq:12,eq:13} and can be calculated with a
quantum chemistry package such as PySCF~\cite{sun_2018_pyscf}. With an initial guess for
$\vec{t}$, we construct a quantum circuit that implements the unitary
$U_{\rm ansatz}(\vec{t})$. The qubits are initialized in the reference state
$\ket{\Phi_{0}}$ and the quantum circuit then prepares $\ket{\Psi(\vec{t})}$. The energy of
this state is found via repeated preparation of state $\ket{\Psi(\vec{t})}$ followed by local
measurements. The energy is given by
\begin{equation}
  \label{eq:40}
  E(\vec{t}) = \sum_{i} H_{i} \prod_{\oldvec{\sigma}} \braket{\Psi(\vec{t})| \sigma^{\alpha}_{j}|\Psi(\vec{t})}
\end{equation}
The parameters $\vec{t}$ are then optimized to obtain the best estimate of the ground
state energy. This optimization is done via a classical algorithm, such as gradient
descent, and hence \ac{vqe} is a hybrid classical-quantum algorithm. At each step of
optimization, the parameters $\vec{t}$ are suitably changed to $\vec{\bar{t}}$ to prepare
a new state $\ket{\Psi(\vec{\bar{t}})}$ and this process is repeated till the energy $E$
converges to a stable value.

Application of the variational principle requires an ansatz which has large overlap with
the ground state wave function. The \ac{ucc} and \ac{kupccgsd} methods construct their
ansatz in a chemically intuitive way. The preparation of such chemically intuitive ansatz
may be resource intensive depending on the hardware design and connectivity of the quantum
computer. Hardware efficient ansatz are a set of trial wave function which can be quickly
prepared on a given quantum computer. One such form of ansatz were used in
Ref.~\cite{kandala_2017_hardwarevqe} to find the ground state energy of several small
molecules. These ansatz cut down the requirement of elaborate state preparation. However,
they might increase the complexity encountered by the classical optimizer as the ansatz
might not systematically approach towards the true ground state. As an extreme example, a
hardware efficient wave function prepared by random application of gates is close to a
maximally mixed state in the Hilbert space, and classical optimization starting from such
state might be hard due to flat energy surface. A systematic approach to generate such
hardware efficient ansatz might show promising results on \ac{nisq} computers.

%%%%%%%%%%%%%%%%%%%%%%%%%%%%%%%%%%%%%%%%%%%%%%%%%%%%%%%%%%%%%%%%%%%%%%%%%%%%%%%%%
\section{Basis sets}
\label{sec:basis-sets}

In \cref{sec:hf}, the starting basis functions $\{\phi_{i}\}$ form a complete set. The
basis set of the set of all arbitrary wave functions is infinite in size and hence the
coefficient matrix $\mytensor{C}$ is infinite dimensional as well. In theory, one can find
an exact solution with only $n$ basis function; this is the basis set where the chosen
functions $\phi_{i}$ happen to be the solution of HF equations~\eqref{eq:hf}. We also need a
finite size basis set to numerically solve the \ac{hf} equations. A well chosen basis set
can still obtain very accurate results with only a finite number of basis functions.

A common starting point for basis functions are the \ac{ao} of hydrogen-like atom. The
\ac{mo}s are linear combination of these \ac{ao}s. This method is appropriately named
\ac{lcao} method. The exact hydrogen-atom like orbital are represented by Laguerre
polynomials. Inspired from these functions, Slater proposed a basis set where the
functions decay exponentially in distance,
\begin{equation}
  \label{eq:sto}
  \phi_{\rm STO}(r; \alpha) \propto p(r)\e^{-\alpha r}
\end{equation}
where $p(r)$ is a polynomial in $r$ and $\alpha$ is an appropriate scale factor. Such basis
functions are called \ac{sto}. While \ac{sto} accurately describe the shape of the atomic
orbitals of an hydrogen-like atom, they are quite cumbersome to use in numerical
integration. It is often helpful to use a basis set where the functions decay in
exponential of the square of the distances,
\begin{equation}
  \label{eq:gto}
  \phi_{\rm GTO}(r; \alpha) \propto p(r) \e^{-\alpha r^{2}} \ .
\end{equation}
These basis functions are called \ac{gto}, since they look similar to a Gaussian
function. Since multiplication of two \ac{gto}s is another \ac{gto}, integrals with these
functions can be simplified significantly. As a down-side, these functions no longer
describe the shape of hydrogen-like orbitals. In order to rectify this problem, basis sets
often use a linear combination of \ac{gto}s to represent a single \ac{ao}. A common basis
set, STO-$n$G, employs a combination to $n$--\ac{gto}s to represent a single \ac{sto};
\begin{equation}
  \label{eq:5}
  \phi_{\rm STO}(r) \approx \phi_{\rm STO-nG} = \sum_{i=1}^{n} c_{i}\, \phi_{\rm GTO}(r; \alpha_{i})
\end{equation}
where the parameters $\{\alpha_{i},c_{i}\}$ are optimized by maximizing the overlap between
exact $\phi_{\rm STO}$ and the approximate $\phi_{\rm STO-nG}$.

The STO-$n$G sets~\cite{hehre_1969_ngto} are a minimal basis set, that is, they only use
one function to represent an atomic orbital. For example, a carbon atom with five orbitals
(1s, 2s, 2${\rm p}_{\rm x}$, 2${\rm p}_{\rm y}$, 2${\rm p}_{\rm z}$) is represented by
five functions. This condition, which is well-reasoned based on the physics of the system,
usually does not result in good numerical results. We can relax this condition to get
better results and use multiple functions to represent the valence orbitals while using a
minimal set for the inner electrons. Such basis set is called a split-valence basis set. A
double-$\zeta$ set such as the such as cc-pVDZ~\cite{dunning_1989_gaussianbasis} basis uses
two functions for each valence orbital, a triple-$\zeta$ uses three, and so on. The numerical
accuracy increases with larger basis sets but more computational resources are required to
determine the larger coefficient matrix $\mytensor{C}$ in \cref{eq:2,eq:hf}.

%%%%%%%%%%%%%%%%%%%%%%%%%%%%%%%%%%%%%%%%%%%%%%%%%%%%%%%%%%%%%%%%%%%%%%%%%%%%%%%%%
\section{Qubit tapering}
\label{sec:qubit-tapering}

Given a problem Hamiltonian with wave function expressed in a basis of $M$ spin orbital
functions ($M/2$ spatial orbital functions), the equivalent quantum circuit requires $M$
qubits, one each to represent an orbital. This number can be reduced by utilizing the
symmetry of the problem Hamiltonian and by using approximations that do not significantly
degrade the quality of the variational solution.

The qubit tapering method~\cite{bravyi_2002_fermionicquantum} tries to reduce the number
of qubits by systematic identification of internal and spatial symmetry of the second
quantized Hamiltonian~\eqref{eq:2quantHam}. Since the Hamiltonian preserves the spin and
the total number of electrons of the system, one can always remove two qubits from their
system via qubit tapering. The method relies on identification of the symmetry generators
of the qubit Hamiltonian. After chosen encoding scheme, the qubit Hamiltonian can be
written as a sum of Pauli strings,
\begin{equation}
  \label{eq:qubit_ham}
  H = \sum_{i} H_{i}\, \prod_{\bm{\sigma}_{i}} \sigma^{\alpha}_{j}
\end{equation}
where $\alpha\in\{x,y,z\}$ specifies one of the Pauli operators and
$\sigma^{\alpha}_{j} \in \bm{\sigma}_{i}$ where $\bm{\sigma}_{i}$ is a set of Pauli operators describing the
Pauli string of $i^{\rm th}$ term. Using techniques adapted from quantum error
correction~\cite{gottesman_1997_stabilizercodes,lidar_brun_qec}, one can find an abelian
group $S$ such that any element of this group commutes with all the Pauli strings of
Hamiltonian~\eqref{eq:qubit_ham}. The size of generator of the symmetry group $S$ is the
number of qubits that can be tapered from Hamiltonian~\eqref{eq:qubit_ham}. In their work,
Bravyi et.\ al.\ were able to remove two aforementioned qubits as well as another qubit in
linear systems such as \ch{H2} and \ch{BeH2}.

We implemented the qubit tapering algorithm for all the dipeptides considered in this
work. We started with a stable conformer of each dipeptide obtained from the PubChem
database~\cite{kim_2019_pubchem2019} and prepared the respective second quantized
Hamiltonian with PySCF~\cite{sun_2018_pyscf}. We encoded the second quantized Hamiltonian
to its qubit form (\cref{eq:qubit_ham}) with \ac{jw} encoding using the Q\#
language~\cite{low_2019_qnwchem}. Finally, we applied the qubit tapering algorithm. We
were not able to eliminate any qubits beyond the aforementioned two qubits from any of the
dipeptide. This result is not surprising since dipeptides are disordered systems and do
not arrange themselves in a symmetric geometry. Nevertheless, such qubit tapering
algorithms might be improved by working on the symmetries of a sub-system of the
Hamiltonian (such as a symmetric aromatic ring which is part of a bigger protein
chain). We leave this question open for future work.

%%%%%%%%%%%%%%%%%%%%%%%%%%%%%%%%%%%%%%%%%%%%%%%%%%%%%%%%%%%%%%%%%%%%%%%%%%%%%%%%%
\section{Common gates}
\label{sec:common-gates}

A quantum algorithm starts with a simple many qubit state, usually the $\ket{0000\ldots0}$
state, and applies various quantum gates which manipulate the qubits accordingly. All
quantum gates can be represented by a unitary operator. We can also interpret these gates
as rotation on the Bloch sphere. Common single qubits gates are the Pauli gates which
rotate the qubits by an angle of $\pi$ along the respective axis:
\begin{equation}
  \label{eq:30}
  \sigma^{x} = \pmat{0 &1 \\ 1 &0} , \quad
  \sigma^{y} = \pmat{0 &-i \\ i &0}, \quad
  \sigma^{z} = \pmat{1 &0 \\ 0 &-1} \ .
\end{equation}
Other common gates are the Hadamard gate $H$, the $Y$ gate and the $T$ gate,
\begin{equation}
  \label{eq:31}
  H = \frac{1}{\sqrt{2}} \pmat{ 1 & 1 \\ 1 & -1}, \
  Y = \frac{1}{\sqrt{2}} \pmat{ 1 & i \\ i &  1}, \
  T = \pmat{ 1 & 0 \\ 0 & \e^{\ii \pi/4}} \ .
\end{equation}
Two qubit gates can be used to generate entangled pair of qubits. The
\ac{cnot} gate is a two qubit gate which applies the $\sigma^{x}$ gate to the second (target)
qubit only if the first (control) qubit is in state $\ket{1}$,
\begin{equation}
  \label{eq:32}
  \mathrm{CNOT} = \ketbra{0}{0}\otimes I + \ketbra{1}{1}\otimes\sigma^{x}
\end{equation}
where $I = \mathrm{diag}(1,1)$ is the single qubit identity matrix. The \ac{cnot} gate
along with the above mentioned single qubit gates form a universal set of quantum gates,
that is, any quantum algorithm (or equivalently, a $n$-qubit unitary) can be decomposed
into a chain of one and two qubit gates (or equivalently, a tensor product of $2\times2$ and
$4\times4$ matrices).

% Balance columns
\makeatletter
\close@column@grid
\twocolumngrid
\makeatother

\end{document}